\documentclass[oldversion]{aa}

\usepackage{graphicx}

\usepackage{txfonts}

\usepackage{color}

\usepackage{natbib}

\usepackage{color}

\def\sss{\scriptscriptstyle}
\def\U{{\sss \!U}}
\def\L{{\sss \!L}}
\def\K{{\sss \!K}}

\def\nur{\nu_\mathrm{r}}
\def\nuv{\nu_\theta}
\def\nuL{\nu_\L}
\def\nuU{\nu_\U}
\def\nuK{\nu_\K}

\begin{document}

\title{
Black hole spin inferred from 3:2 epicyclic resonance model of high-frequency quasi-periodic oscillations
}

\author
{E. \v{S}r\'amkov\'a\inst{\nabla}, G. T\"or\"ok\inst{\nabla}, A. Kotrlov\'a\inst{\nabla}, P. Bakala\inst{\nabla}, M. A. Abramowicz\inst{\nabla}\inst{\Delta}, Z. Stuchl\'{\i}k\inst{\nabla}, K. Goluchov\'a\inst{\nabla}\inst{\Delta}, W.~Klu{\'z}niak\inst{\nabla}\inst{\Delta}}

\institute{$\nabla$ Institute of Physics, Faculty of Philosophy and Science, Silesian
  University in Opava, Bezru\v{c}ovo n\'{a}m. 13, CZ-74601 Opava, Czech Republic\\
$\Delta$ Nicolaus Copernicus Astronomical Centre, Bartycka 18, PL-00716 Warsaw, Poland}
  
\date{Received / Accepted}

\authorrunning{E. \v{S}r\'amkov\'a et al.}
\titlerunning{Black hole spin inferred from 3:2 resonance model}
 
\date{Received / Accepted}

\abstract
{Estimations of black hole spin in the three Galactic microquasars GRS 1915+105, GRO J1655-40, and XTE J1550-564 have been carried out based on spectral and timing X-ray measurements and various theoretical concepts. Among others, a non-linear resonance between axisymmetric epicyclic oscillation modes of an accretion disc around a Kerr black hole has been considered as a model for the observed high-frequency quasi-periodic oscillations (HF QPOs). Estimates of spin predicted by this model have been derived based on the geodesic approximation of the accreted fluid motion. Here we assume accretion flow described by the model of a pressure-supported torus and carry out related corrections to the mass-spin estimates. We find that for dimensionless black hole spin $a\equiv cJ/GM^2\lesssim0.9$, the resonant eigenfrequencies are very close to those calculated for the geodesic motion. Their values slightly grow with increasing torus thickness. These findings agree well with results  of 
 a previous study carried out in the pseudo-Newtonian approximation. The situation becomes different for $a\gtrsim0.9$, in which case the resonant eigenfrequencies rapidly decrease as the torus thickness increases. We conclude that the assumed non-geodesic effects shift the lower limit of the spin, implied for the three microquasars by the epicyclic model and independently measured masses, from $a\sim0.7$ to $a\sim0.6$. Their consideration furthermore confirms compatibility of the model with the rapid spin of GRS~1915+105 and provides highly testable predictions of the  QPO frequencies. Individual sources with a moderate spin ($a\lesssim0.9$) should exhibit a smaller spread of the measured 3:2 QPO frequencies than sources with a near-extreme spin ($a\sim1$). This should be further examined using the large amount of high-resolution data expected to become available with the next generation of X-ray instruments, such as the proposed Large Observatory for X-ray Timing (LOFT).}
\maketitle

\section{Introduction}
\label{section:introduction}

\begin{table*}
\caption{Properties of the three microquasars GRO 1655-40, GRS 1915+105, and XTE 1550-564. The individual columns display the frequencies of the lower and upper QPO peaks \citep{str:2001a,str:2001b,rem-etal:2002,rem-etal:2003}, the mass estimates \citep{gre-etal:2001, grn-etal:2001, oro-etal:2002, mcc-rem:2003,mcc-rem:2006}, and the spin values inferred from the epicyclic resonance model assuming a geodesic accretion flow.\label{table:1}
}
\centering
\renewcommand{\arraystretch}{1.4}
\vspace{-3ex}
\begin{tabular}{lcccc}\\ \hline \hline
Source & $\nu_{\mathrm{L}}$ [Hz] & $\nu_{\mathrm{U}}$ [Hz] & Mass [$\mathrm{M}_{\odot}$] & $a$ \\ \hline
GRO 1655-40 & 300 &  450 & 6.0--6.6\,\tablefootmark{a} & 0.96--0.99\\
\hline
GRS 1915+105 & 113 &  168 & 10.0--18.0\,\tablefootmark{b} & 0.68--0.99\\
\hline
XTE 1550-564 & 184 &  276 & 8.4--10.8 &  0.89--0.99\\
\hline
\end{tabular}
\tablefoot{
\tablefoottext{a}{Alternative predictions of the black hole mass in GRO~1655-40 are $5.1 - 5.7\,\mathrm{M}_{\odot}$ \citep{Bee-Pod:2002:} and $5.8 - 6.8\,\mathrm{M}_{\odot}$ \citep{gre-etal:2001}.
\tablefoottext{b}{For GRS 1915+105, \citet{Rei-etal:2014:} have recently obtained a tighter mass estimate, $10.6 - 14.4\,\mathrm{M}_{\odot}$, based on a trigonometric parallax measurement of the distance \citep[see also][]{stee-etal:2013}.}
}}
\end{table*}
 
Studying the X-ray spectra and variability provides a useful tool for probing strong-field gravity effects and places constraints on the properties of compact objects, such as the mass or spin of a black hole. One of the commonly accepted ways to measure the black hole spin is related to fitting the X-ray spectral continuum or the relativistically broadened Fe K alpha lines. Using the spectral fitting methods, estimations of black hole spin have been carried out by many authors in the past \citep[see e.g.,][]{mcc-etal:2006,mcc-etal:2008,mid-etal:2006,don-etal:2007,mil:2007,sha-etal:2008,mcc-etal:2010,mcc-etal:2011,mcc-etal:2014}. 

There is another approach to estimating parameters of accreting black holes (BH) and neutron stars (NS) that has been gaining popularity in the past decade. It is the determination of compact object properties through the theory of high-frequency quasi-periodic oscillations (HF QPOs). These peaked features have been observed in the  X-ray power density spectra of the BH and NS low-mass X-ray binaries for several decades \citep[see e.g.,][for a review]{Kli:2006:CompStelX-Ray:,bel-ste:2014}. In the BH systems, the HF QPOs appear at frequencies that often form rational ratios with a preferred ratio of 3:2 \citep[][]{abr-klu:2001,mcc-rem:2003}.
\bigskip

\subsection{Orbital models of HF QPOs}

A rich variety of models has been proposed to explain the BH HF QPOs. Many of them are associated with motion of matter accreting onto the central compact object \citep[e.g.,][and others]{kat-fuk:1980,alp-sha:1985,lam-etal:1985,oka-etal:1987,now-wag:1991,per-etal:1997,klu:1998,mil-etal:1998,psa-etal:1999,ste-vie:1998a,ste-vie:1998b,wag:1999,wag-etal:2001,sil-etal:2001,abr-klu:2001,tit-ken:2002,rez-etal:2003,pet:2005,zha:2005,sra-etal:2007,kat:2007,cad-etal:2008,stu-etal:2008,hor-etal:2009,muk:2009,lai-etal:2012,ort-etal:2014}. 
Despite the different involved physical mechanisms, several of the HF QPO models assign the observed frequencies to  orbital
frequencies that can under some approximations be expressed in terms of Keplerian and epicyclic frequencies of perturbed circular geodesic motion.

One of the early models, for instance, the so-called relativistic-precession model \citep{ste-vie:1998a,ste-vie:1998b,ste-vie:1999}, identifies the kHz QPO frequencies $\nuU$ and $\nuL$  with the Keplerian orbital frequency and the periastron-precession frequency, 
$
\nuU=\nuK,~\nuL=\nuK - \nur
$
 (where $\nu_r$ denotes the radial epicyclic frequency). Within the model framework, it is usually assumed that the observed kHz variability originates in a bright localized spot or blob orbiting the black hole at a slightly eccentric orbit. The photons emitted by the spot, which have a high orbital velocity of the order of percents of the speed of light, are then propagated from the strong gravity region towards the observer. Consequently, the detected X-ray flux is periodically modulated by the relativistic effects \citep[see e.g., ][and references therein]{sch-ber:2004,bak-etal:2014}. 

A different class of models is based on the assumption that the light curve reflects variations in the emission of an accretion disk. A theory of normal modes of accretion disk oscillation, referred to as discoseismology, has been developed in the papers coauthored by Kato and by Wagoner. For instance, \citet{kat-fuk:1980} consider the g-mode of disk oscillation, whose frequency is close to the highest radial epicyclic frequency, while \citet{sil-etal:2001} developed a theory of c-mode (disc corrugation) whose frequency is related to the difference between the vertical epicyclic frequency and the orbital frequency.

\subsubsection{Resonance models}

\cite{klu-abr:2001} have introduced another concept of orbital models. In these models, the relativistic effects acting on the emitted photons are also essential, but the HF QPOs are attributed to a non-linear resonance between two modes of accretion disc oscillations. This well explains the observed preferred frequency ratio, $\nuU/\nuL = 3/2$. Since the concept deals with a collective motion of accreted matter, it has a high potential of the observed flux modulation \citep[][]{bur-etal:2004,maz-etal:2013,vin-etal:2014,bak-etal:2014}.

The framework of resonance models has been extensively
developed \citep[][]{abr-klu:2001,abr-etal:2003, klu-etal:2004, hor-kar:2006, hor-etal:2009} and miscellaneous combinations of disc oscillation modes have been considered. In several particular cases, the observed frequencies can be expressed in terms of Keplerian and epicyclic frequencies \citep[see][for details and references]{tor-etal:2005,hor:2008,klu:2008}.

\subsection{Spin estimation and epicyclic resonance model}

In Kerr geometry, which describes the spacetime around a rotating black hole, the frequencies of the radial and vertical epicyclic oscillations depend on the mass and spin of the black hole. It is therefore possible to determine the BH mass or spin from the epicyclic resonance  
model and the observed 3:2 QPO frequencies. For three Galactic microquasars with the 3:2 HF QPOs (GRS 1915+105, GRO J1655-40, and XTE J1550-564), such BH spin estimations have been carried out in the past for the epicyclic resonance model and for several other HF QPO models \citep[][]{str:2001a,str:2001b,wag-etal:2001,abr-klu:2001,kat:2004a,kat:2004b,tor-etal:2005,tor-etal:2011,mot-etal:2014a,mot-etal:2014b}. Analogous spin estimations have been obtained for the three sources through other methods by different groups of authors using the X-ray continuum-fitting procedure \citep[][]{mcc-etal:2006,mcc-etal:2008,mid-etal:2006,mil-etal:2009}, and the iron line method \cite[][]{mil:2007,dav-etal:2006,mcc-rem:2009,blu-etal:2009,ste-etal:2011}. The spin estimates obtained through the spectral and timing approaches are as yet somewhat inconsistent, however \citep[see a detailed discussion in][]{tor-etal:2011}.

In this work, we consider one of the popular {versions} of resonance models represented by the epicyclic parametric (or internal) resonance model that assumes a 3:2 non-linear resonance between the axisymmetric radial and vertical epicyclic modes of accretion disc oscillations. All the estimates derived so
far that were related to the epicyclic resonance model assume
a geodesic approximation of the accreted fluid motion. Within this approximation, the two observable resonant frequencies are represented by the radial and vertical epicyclic frequencies, $\nur$ and $\nuv$, {of  test} particle motion. In more general accretion flows, non-geodesic effects connected for instance
to pressure gradients, magnetic fields, or other forces may affect the properties of the considered oscillation modes and consequently the inferred spin predictions.

In the following, we explore black hole spin predictions implied by the epicyclic resonance QPO model considering non-geodesic influence that originates in pressure forces present in accretion flow, which is modelled by a small equilibrium-pressure-supported, perfect-fluid torus. Properties of the epicyclic modes of torus oscillations, such as modifications to their frequencies due to pressure gradients present in the torus, were calculated by \cite{bla-etal:2007} in the pseudo-Newtonian approximation and were later generalised by \cite{str-sra:2009} for Kerr geometry. Using the results of \cite{str-sra:2009}, we apply the modified epicyclic frequencies to calculate the corresponding estimates of black hole spin based on the epicyclic resonance QPO model.

\section{Resonant eigenfrequencies and black hole parameters implied by the epicyclic model}
\label{section:spin}

The explicit formulae of the epicyclic frequencies in Kerr geometry were first derived by \cite{ali-gal:1981} and may be written in the Boyer-Lindquist coordinates $t,r,\theta,\phi$ as \citep[e.g.,][]{sil-etal:2001,tor-stu:2005}
\smallskip
\begin{eqnarray}
\label{radial}
\nu_r^2 &=&\alpha_r\,\nu_\mathrm{K}^2,
\\
\label{vertical}
\nu_{\theta}^2 &=&\alpha_\theta\,\nu_\mathrm{K}^2,
\end{eqnarray}
where 
\begin{eqnarray}
\label{Keplerian}
&&\nu_{\mathrm{K}}=\left ({{GM}\over {r_{\rm G}^{~3}}}\right )^{1/2}\left( x^{3/2} + a \right)^{-1},
\\
\nonumber
\\
&&\alpha_r\left( x\,,a\right)\equiv{1-6\,x^{-1}+ 8 \,a \, x^{-3/2} -3 \, a^2 \, x^{-2}},
\\
&&\alpha_\theta\left( x\,,a\right)\equiv{1-4\,a\,x^{-3/2}+3a^2\,x^{-2}},
\\
&&x=r/r_{\rm G}, ~~~ r_{\rm G}=GM/c^2.
\end{eqnarray}
Here $M$ and $a$ denote the mass and the dimensionless angular momentum (spin) of the black hole.

Within the {epicyclic resonance} (ER) model, it is assumed that the resonant eigenfrequencies are equal to epicyclic frequencies defined at the $x_{3:2}$ orbit where we have
\begin{equation}
\nuv=3/2\nur\,.
\end{equation}
Furthermore, it is assumed that the eigenfrequencies are equal to the observed upper and lower 3:2 QPO frequencies, $\nuU$ and $\nuL$ \citep[e.g.,][]{tor-etal:2005}. Using this set of formulae and the observed 3:2 HF QPO frequencies along with the independently estimated black hole masses, the intervals of spin predicted by the ER model were previously calculated for the three Galactic microquasars \citep[GRS 1915+105, GRO J1655-40, and XTE J1550-564 -- see][]{tor-etal:2005,tor-etal:2011}. The obtained results and relevant properties of the three sources are summarized in Table~\ref{table:1}.

\subsection{Epicyclic frequencies in a slightly non-slender torus}

\begin{figure}[t]
\begin{center}
\includegraphics[width=.75\hsize]{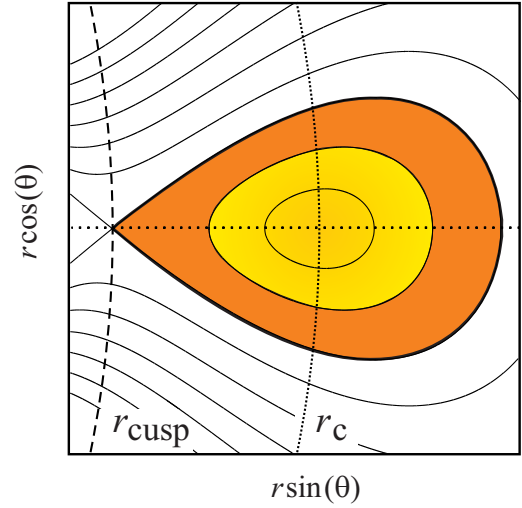}
\end{center}
\caption{Illustration of equipotential surfaces of the accretion torus. The dotted horizontal line denotes the equatorial plane. The yellow region illustrates tori of various thickness. The orange region, together with the yellow, illustrates a torus filling the equipotential surface with a cusp.}
\label{figure:cusp}
\end{figure}

We assumed a non-geodesic flow that was modelled by an equilibrium, slightly non-slender, pressure-supported perfect-fluid torus. The torus was assumed to have a constant specific angular momentum distribution and orbit a rotating Kerr black hole. We assumed that the torus is formed by a perfect fluid that fills up closed equipotential surfaces. One of these surfaces has a cusp in the equatorial plane, which embodies features similar to those of the L1 Lagrange point. The topology of the equipotential surfaces and the equilibrium torus is illustrated in Fig.~\ref{figure:cusp}.\footnote{For a detailed description of the model of the torus and further references see \cite{str-sra:2009}.}.

In the investigated non-geodesic flow, the radial and vertical epicyclic oscillations of the fluid are modified by pressure forces. Using a perturbation theory, within the accuracy of a second order in the thickness of the torus, \cite{str-sra:2009} calculated explicit formulae for the pressure corrections to epicyclic frequencies in a slightly non-slender, constant specific angular momentum torus orbiting a Kerr black hole. The calculated epicyclic frequencies can be written in the following form:
\begin{eqnarray}
\label{frequencies-r}
\nu_{\rm r}^* &=& \nu_r + \beta^2C_r(M,r_{\rm{c}},a), ~~~ C_r(M,r_{\rm{c}},a)<0,
\\\,
\label{frequencies-t}
\nu_{\rm \theta}^* &=& \nu_{\theta} + \beta^2C_\theta(M,r_{\rm{c}},a), ~~~ C_\theta(M,r_{\rm{c}},a)<0,
\end{eqnarray}
where $C_r(M,r_{\rm{c}},a)$ and $C_\theta(M,r_{\rm{c}},a)$ denote the negative pressure corrections evaluated at the centre of the torus, $r=r_{\rm{c}}$.\
Explicit forms of expressions for the epicyclic frequencies (\ref{frequencies-r}) and (\ref{frequencies-t}) are given in formulae (52) and (56) of \cite{str-sra:2009}.

\begin{figure*}[h!t]
\begin{center}
a) \hfill b) \hfill~\\
\includegraphics[width=1\hsize]{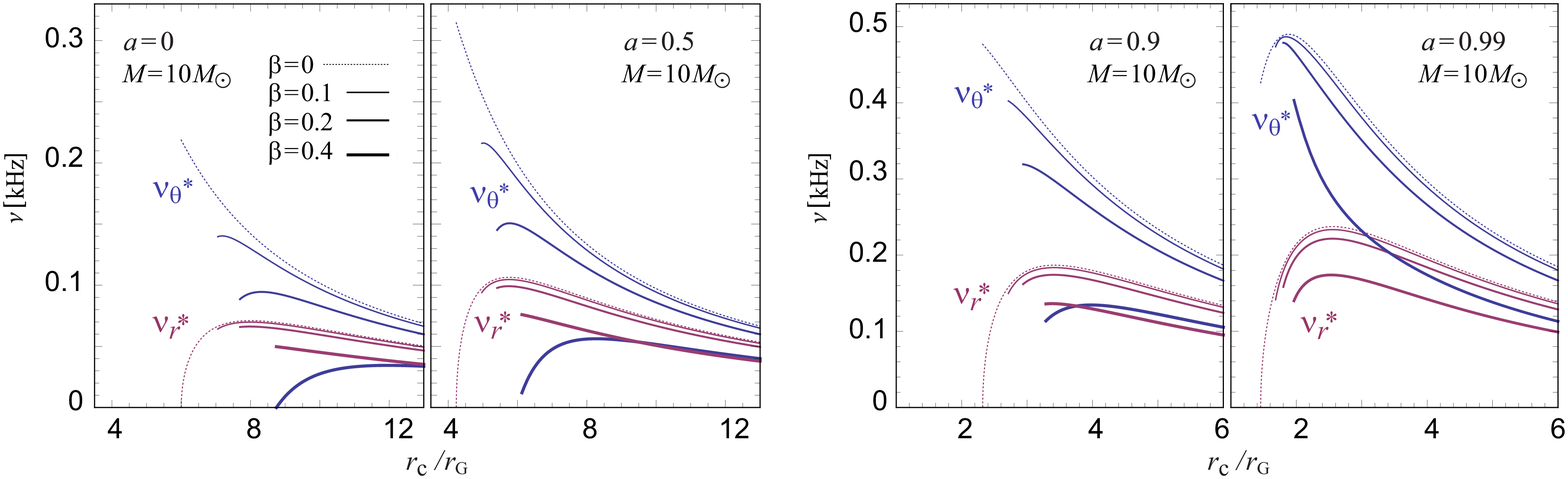}
\end{center}
\caption{Epicyclic frequencies calculated at the centre of the torus, $r=r_{\rm{c}}$, plotted for various torus thicknesses ($\beta$) and black hole spins $a$. The curves are drawn solely within the region where the considered value of $\beta$ does not exceed the critical value corresponding to a torus with a cusp.}
\label{figure:1}
\end{figure*}

The dimensionless parameter $\beta$ determines the torus thickness. It is defined as \citep{bla:1985}
\begin{equation}
\label{beta}
\beta^2 \equiv \frac{2nc_{\rm s}^2} {r_{\rm{c}}^2 \nu_{\rm{Kc}}^2 U^2},
\end{equation}
where $c_{\rm s}$, $\nu_{\rm{Kc}}$, and $U$ are the sound speed, Keplerian frequency, and $t$-component of the contravariant four-velocity of the perfect fluid, all defined at the centre of the torus, $r=r_{\rm{c}}$. Within the perturbative approach applied by \cite{str-sra:2009}, it is expected that $0\leq\beta^2\ll1$. {For the critical $\beta$ values corresponding to tori with a cusp, the limit of $\beta\rightarrow 0$ corresponds to an infinitely slender torus located at the marginally stable circular orbit.} High values of $\beta<1$ then correspond to very large tori whose inner edge approaches the location of the marginally bound circular orbit. The behaviour of epicyclic frequencies for tori of different thickness, $\beta \in  \left\{0,~0.1,~0.2,~0.4\right\}$, and a Kerr black hole with $a\in\left\{0.0,~0.5,~0.9,~0.99\right\}$ is illustrated in Fig.~\ref{figure:1}.

\subsection{Behaviour of commensurable frequencies for moderately and rapidly rotating Kerr black holes}

\begin{figure*}[th!]
\begin{center}
a) \hfill b) \hfill c) \hfill d) \hfill~\\
\includegraphics[width=1.\hsize]{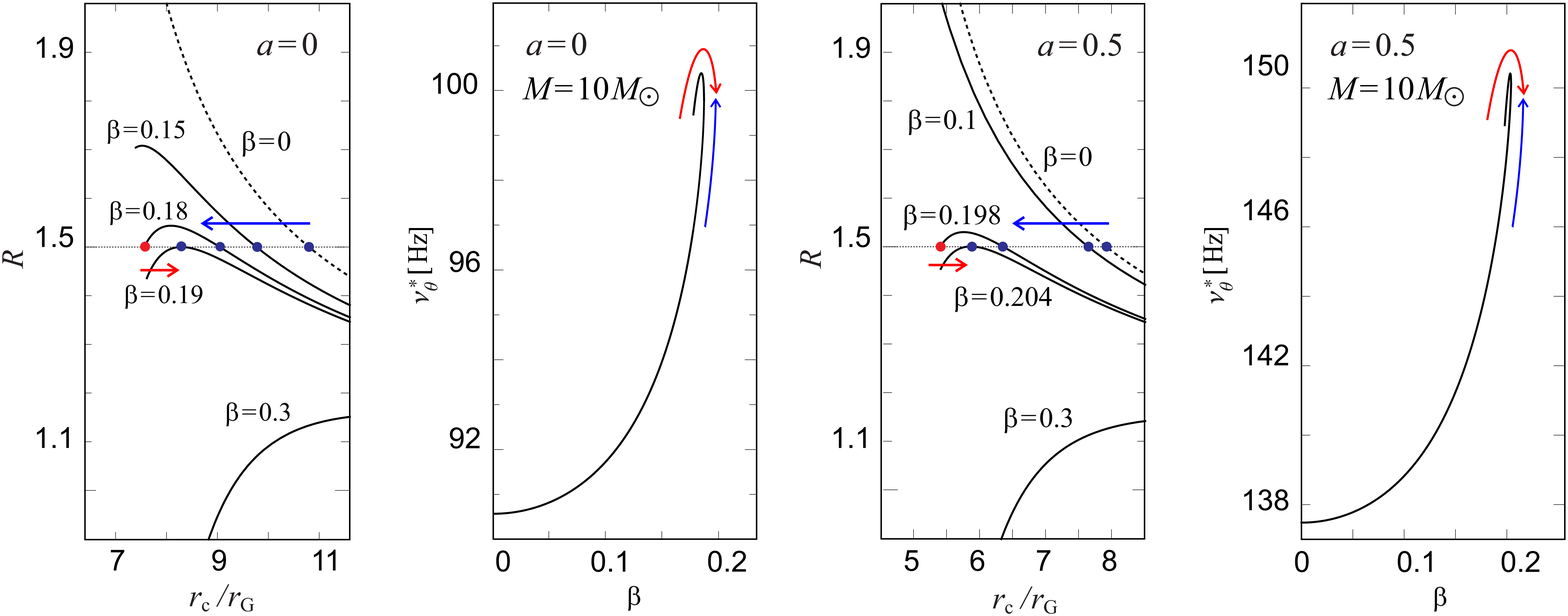}
\\
e) \hfill f) \hfill g) \hfill h) \hfill~\\
\includegraphics[width=1.\hsize]{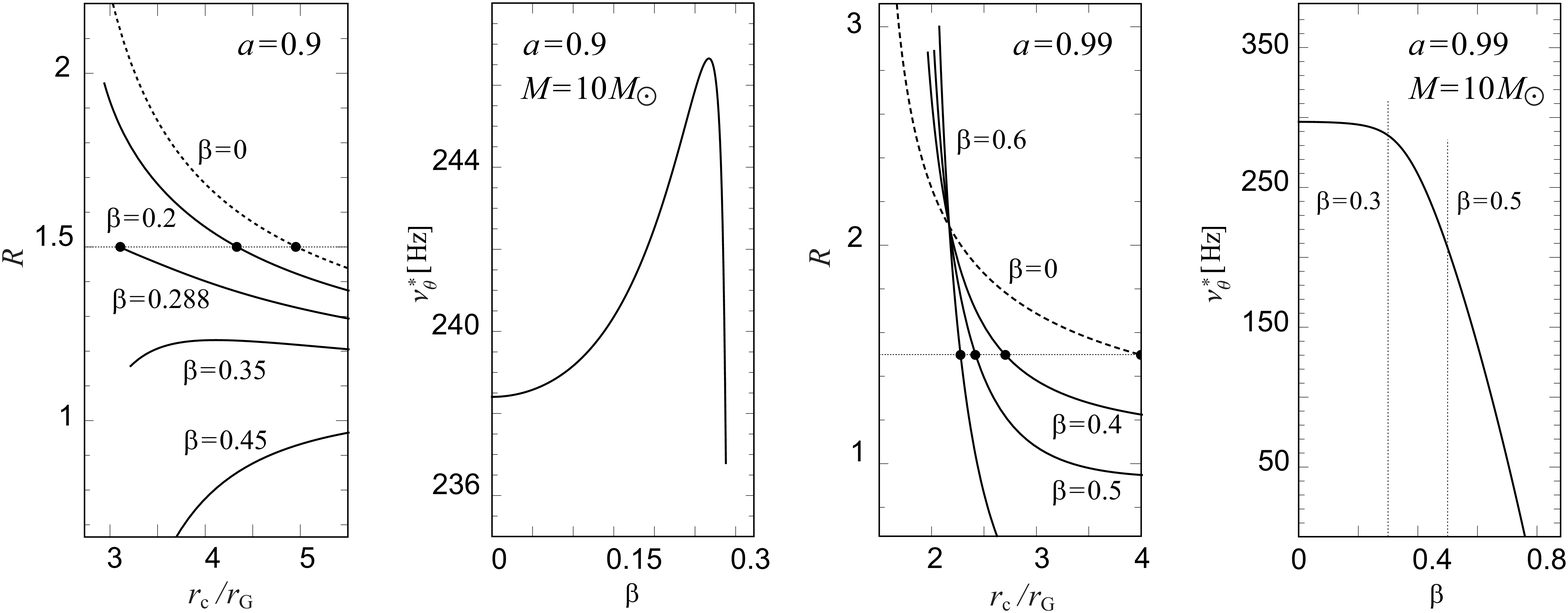}
\end{center}
\caption{Ratio $R$ of the vertical to radial epicyclic frequency and the resonant vertical epicyclic frequency $\nu_{\rm \theta}^{\,*}$ calculated at the radius $x_{3:2}$. The arrows in panels a)-d) indicate the behaviour corresponding to the two resonant radii $x_{3:2}$. The inner (red) and outer (blue) resonant radius is denoted.}
\label{figure:2}
\end{figure*}

Using formulae (\ref{frequencies-r}) and (\ref{frequencies-t}), we investigated the behaviour of epicyclic frequencies in fluid tori when they are in the 3:2 ratio. For Schwarzschild and moderately rotating Kerr black holes, the resonant frequencies behave qualitatively in the same fashion. The resonant radius in this case decreases with growing torus thickness, while the individual epicyclic frequencies defined at this radius increase with growing torus size. This is illustrated in Fig. \ref{figure:2}, which, for different values of $a$ and $\beta$, displays the ratio $R$ of the vertical to radial epicyclic frequency, and the vertical epicyclic frequency calculated at the resonant radius. Panels a) and c) of Fig. \ref{figure:2} drawn for $a=0$ and $a=0.5$ show that from certain value of $\beta$, there suddenly appear two different tori whose epicyclic frequencies display the 3:2 commensurability. This behaviour persists as we continue to increase the torus size until at some point the 3:2  resonant radius no longer exists. This phenomenon manifests itself in the form of a small loop in the profile of the resonant vertical epicyclic frequency plotted over $\beta$ (panels b) and d) of Fig. \ref{figure:2}). The same behaviour was found by \cite{bla-etal:2007}, who carried out a similar analysis in the pseudo-Newtonian approximation. We find that for Kerr black holes, two different values of resonant frequency are allowed for the same value of $\beta$ when $a\leq0.86$.

For higher values of $a$, the behavioural nature of the resonant frequencies begins to change significantly. The resonant radius retains its decreasing pattern with growing torus size, but two different tori with the 3:2 commensurability are no longer possible. Rather distinct modifications also appear in the character of the resonant frequency, which at first increases with rising $\beta$, but at some point starts to decrease rapidly. For a
black hole spin approaching $a\doteq 0.89$, the resonant frequencies implied by the highest allowed values of $\beta$ become lower than those calculated for the geodesic motion. This is illustrated in panels e) and f) of Fig. \ref{figure:2}. For a still higher black hole spin ($a\gtrsim0.99$), characteristic features of the resonant frequencies continue to change even further. While the resonant radius continues to drop monotonically with growing torus thickness, resonant frequencies are no longer higher than those of free test particles. They now decrease for tori of all sizes ($\beta>0$), see panels g) and h) of Fig. \ref{figure:2}. This is because the nature of epicyclic frequencies for rapidly rotating black holes is such that the decrease of frequencies due to higher $\beta$ overrides the increase implied by lower resonant radius.

\subsection{Comparison of predicted 3:2 epicyclic frequencies with observation}
\label{section:intervals}

\begin{figure}[ht!]
\begin{center}
\includegraphics[width=0.95\hsize]{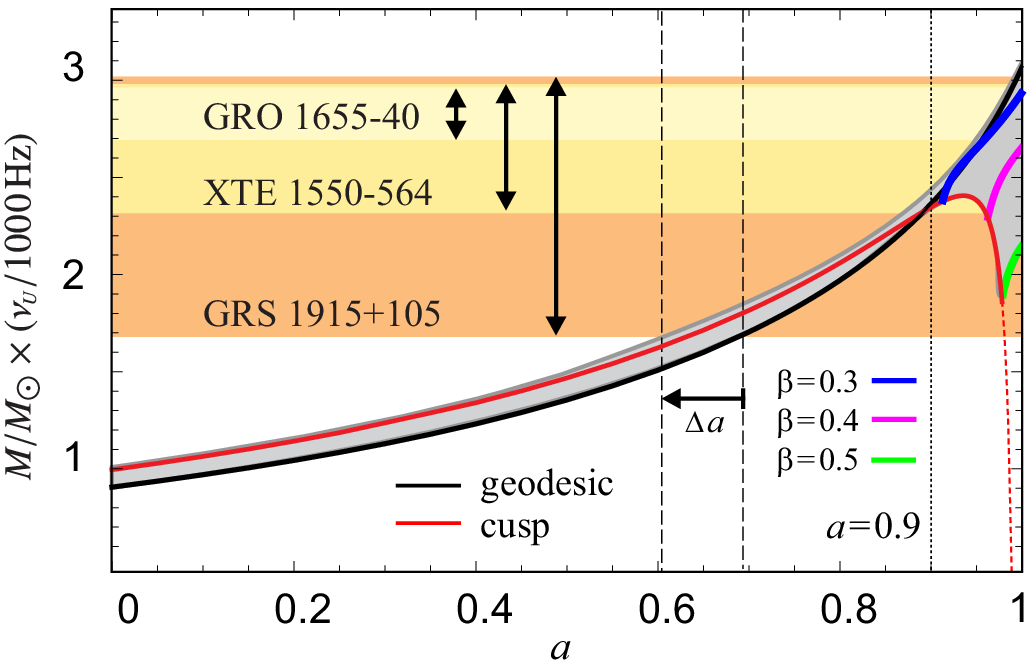}
\end{center}
\caption{Observational bounds on the quantity $M\times\nuU$ obtained for Galactic microquasars vs. relations predicted by the epicyclic resonance model. The vertical areas denote the observational bounds on $M\times\nuU$ determined for GRS~1915$+$105, XTE~J1550$-$564, and GRO~J1655$-$40. The areas are drawn assuming the conservative limits from Table~\ref{table:1}. The solid black curve shows predictions of the model calculated for the geodesic flow. The shaded region indicates the interval predicted by the model for the family of equilibrium non-slender fluid tori of all thicknesses possible in the particular case. The red curve corresponds to the case of tori with a cusp. The arrow labelled $\Delta\,a$ indicates the shift of the lower limit on the spin of microquasars implied by considering the non-geodesic flow.}
\label{figure:3}
\end{figure}

Figure \ref{figure:3} uses the relativistic $1/M$ scaling of the orbital frequencies. It displays the $M\times\nuU(a)$ relation expected from the geodesic epicyclic resonance model marked by the thick black line. The shaded region describes the mass-spin interval predicted by the epicyclic resonance model for a slightly non-slender torus of various thickness. In the figure, we also present the conservative observational bounds on $M\times\nuU$ determined for the three microquasars (see Table~\ref{table:1}). These bounds are indicated by the colour-coded areas. Apparently, while the geodesic line enters the observationally determined vertical range of microquasars when $a\gtrsim 0.7$, the non-geodesic line enters this range already for $a\gtrsim0.6$.\footnote{According
to individual studies based on spectral and timing methods, it would also be possible to include several (vertical) bounds on the spin of microquasars. Within the figure, however, we focus primarily on the epicyclic resonance model of HF QPOs and its predictions. The discussion of various spectral and timing estimates of $a$ can be found for instance in 
\cite{mil:2007}, \citeauthor{mcc-etal:2011} (\citeyear{mcc-etal:2011}, \citeyear{mcc-etal:2014}), \cite{ste-etal:2013}, \cite{tor-etal:2011}, and \cite{kot-etal:2014}.}

Figure \ref{figure:3} clearly illustrates that for $a\lesssim0.9$, the highest possible increase of the predicted resonant frequency due to pressure effects is rather small, reaching about $10\%$. However, for $a\gtrsim 0.9$, the shaded region in the figure shifts from the area above the $\beta=0$ curve to the area located mostly below this curve. The strongest possible shift of the resonant frequency is now much greater than it was for $a \lesssim 0.9$. The coloured curves in Fig.~\ref{figure:3} illustrate the mass-spin values predicted by the model for tori of $\beta=0.3$, $\beta=0.4$, and $\beta=0.5$.  In this case, predictions of the model with non-geodesic flow may significantly differ from those calculated for the geodesic flow. This has a direct link to observations, since the calculated shaded region overlaps a large part of the observationally determined yellow area.  


\section{Discussion and conclusions}
\label{section:conclusions}

Following the previous studies of \cite{bla-etal:2007}, \cite{str-sra:2009}, and \cite{tor-etal:2011},  we explored here the implications of consideration of pressure forces that originate in a pressure-supported fluid torus for the predictions of black hole spin inferred from the epicyclic resonance QPO model. Our findings for moderately rotating Kerr black holes with $a\lesssim 0.9$ {agree well} with those carried out in the pseudo-Newtonian study of \cite{bla-etal:2007}. The epicyclic resonance model with a non-geodesic flow predicts the black hole spin to be slightly lower than the spin estimated previously by the same model with a geodesic flow. The lower limit on the spin of Galactic microquasars inferred by the model is then decreased to $a\sim0.6$. These findings indicate that for up to $a\sim0.9$, the model has a high predictive power (falsifiability).

Different but very interesting results appear for rapidly rotating black holes with $a\gtrsim 0.9$. In this case, the behaviour of the resonant frequencies begins to markedly change in comparison to the pseudo-Newtonian case. This occurs in such a way that the predicted frequencies can be both higher and lower than those predicted by the geodesic model, with the strongest possible shift being much more significant towards the lower values. Therefore, when the source mass is determined from the 3:2 QPO frequencies, 
the epicyclic resonance model has a lower predictive power for rapidly rotating black holes than for moderately rotating black holes. On the other hand, our findings confirm the compatibility of the model with the rapid spin of the microquasar GRS~1915+105 \citep[][]{mcc-etal:2006,tor-etal:2011,fra-mcc:2014}, since the expected mass interval almost overlaps with the interval constrained observationally. Furthermore, we provided highly testable predictions of the QPO frequencies. When the torus size ($\beta$) is not determined by some additional physical requirements, a wide range of commensurable frequencies is predicted for near-extreme rotating black holes.\footnote{Values of $\beta$ as high as $\beta=0.5$ considered in panel h) of Fig.~\ref{figure:2} may require additional treatment, since these values exceed the limits of full applicability of the adopted approximation. Nevertheless, the illustrated phenomenon is very significant and well apparent already for $\beta\sim0.3
 $.} Individual sources with a moderate spin ($a\lesssim0.9$) should therefore exhibit a smaller spread of the measured 3:2 QPO frequencies than sources with a near-extreme spin ($a\sim1$). At present,  there is a rather small amount of data needed for further analysis of this issue. Our predictions, however, can be explored using the large amount of high time-resolution data expected to become available with the next generation of X-ray instruments, such as the proposed Large Observatory for X-ray Timing \citep[LOFT;][]{fer-etal:2012}.

\section*{Acknowledgments}
{E\v{S} and GT would like to acknowledge the Czech Science Foundation grant GA\v{C}R 209/12/P740. ZS acknowledges the Albert Einstein Centre for Gravitation and Astrophysics supported by the Czech Science Foundation grant No. {14-37086G}. Furthermore, we acknowledge financial support from the internal grants of SU Opava, SGS/11/2013, SGS/23/2013, and IGS/9/2015.}



\end{document}